\documentclass[aps,pra,numerical,reprint,onecolumn,showpacs]{revtex4-1}  
\usepackage[dvips]{graphicx}%
\usepackage{bm}%
\usepackage{multirow}
\usepackage{amsmath}
\usepackage{amssymb}
\usepackage{MnSymbol}
\usepackage{dcolumn}
\usepackage[normalem]{ulem} 
\usepackage[colorlinks=true,linkcolor=blue,citecolor=blue,urlcolor=blue]{hyperref}

\begin{document}

\newcommand{\op}[1]{{\bm{#1}}}
\newcommand{\bra}{\langle}
\newcommand{\ket}{\rangle}
\newcommand{\new}[1]{\textcolor[rgb]{0,0,0.7}{#1}}
\newcommand{\old}[1]{\textcolor[rgb]{1,0,0}{\sout{#1}}}
\newcommand{\nota}[1]{\textcolor[rgb]{1,0,0}{{#1}}}
\newcommand{\Oplus}{\ensuremath{\vcenter{\hbox{\scalebox{1.5}{$\oplus$}}}}}

\renewcommand{\multirowsetup}{\centering}
\newlength{\LL}\settowidth{\LL}{Atomic Levels} 

\title{Effect of the Atomic Dipole-Dipole Interaction on the Phase Diagrams of Field-Matter Interactions I: Variational procedure}
\author{S. Cordero} 
\email{sergio.cordero@nucleares.unam.mx}
\author{O. Casta\~nos}
%
\author{R. L\'opez-Pe\~na}
%
\author{E. Nahmad-Achar}
%

%
\affiliation{
Instituto de Ciencias Nucleares, Universidad Nacional Aut\'onoma de M\'exico, Apartado Postal 70-543, 04510 Cd.Mx., Mexico}


\begin{abstract}
We establish, within the second quantization method, the general dipole-dipole Hamiltonian interaction of a system of $n$-level atoms.   The variational energy surface of the $n$-level atoms interacting with $\ell$-mode fields and under the Van Der Waals forces is calculated with respect the tensorial product of matter and electromagnetic field coherent states.  This is used to determine the quantum phase diagram associated to the ground state of the system and quantify the effect of the dipole-dipole Hamiltonian interaction. By considering real induced electric dipole moments, we find the quantum phase transitions for $2$- and $3$-level atomic systems interacting with $1$- and $2$- modes of the electromagnetic field, respectively.  The corresponding order of the transitions is established by means of Ehrenfest classification; for some undetermined cases, we propose two procedures: the difference of the expectation value of the Casimir operators of the $2$-level subsystems, and by maximizing the Bures distance between neighbor variational solutions.
\end{abstract}

%
%

\maketitle

\section{Introduction}

Recently, we have studied the quantum phase diagrams of a system of $n$-level atoms interacting with $\ell$ electromagnetic modes in a cavity, under the dipolar aproximation~\cite{cordero21, lopez-pena21}.

When the inter-atomic distance of a cold atomic gas is comparable to the wavelength of the electromagnetic field, the dipole-dipole coupling between the atoms becomes important and yields relevant collective effects~\cite{weiner99}. These Van der Waals forces, due to dipole-dipole interactions of the induced electric dipole moments, become important and must be taken into account. However, one needs to be careful about the long or short character of the dipolar potential for many particle systems, as one can find in theoretical and experimental studies of ultra-cold boson systems~\cite{jones06, lahaye09, astrakharchik08}.

The dipole-dipole interaction decays as $1/r^3$, with $r$ the distance between particles, and is thus of a different nature as the matter-field interaction considered in earlier works (cf., e.g.,~\cite{cordero15} and references therein). Energy transfer between the particles (atoms, molecules) is one of the important consequences of this interaction. For Rydberg atoms it is particularly interesting, as they have high principal quantum numbers $n$, while the dipole moment scales as $n^2$ in atomic units~\cite{gerry05}.

A review of theoretical and experimental work on the dipole-dipole interaction between Bose-Einstein condensates has been presented in~\cite{lahaye09}. The trapping of cooled polar molecules and other atomic species~\cite{Doyle04} was important for attracting the attention to study these type of interactions, and the long-range dipole-dipole interaction in low-density atomic vapors was detected in~\cite{yu19}, confirming that the interaction is indeed long-range, and that it is present at any density.
 
According to the previous discussion, the interaction between atoms might be relevant to the determination of the quantum phase diagrams for the system constituted by $n$-level atoms interacting with $\ell$-modes of electromagnetic radiation in a cavity. The main objective of this work is to quantify the effect of the atomic dipole-dipole induced interaction on the properties of the ground state of the system. The original contributions of this work are the following: To establish the general dipole-dipole interaction Hamiltonian for a system of $n$-level atoms interacting with $\ell$-modes of electromagnetic radiation in a cavity. To calculate the associated energy surface, which allows us to determine the variational ground state, playing a fundamental role in finding the quantum phase diagrams of the system. The cases for $2$- and $3$- level atomic configurations are worked out explicitly, determining the quantum phase diagrams together with the corresponding order of the transitions. It is remarkable that, even for a finite number of atoms, the surface of maximum Bures distance is able to detect the phase transitions where the Ehrenfest method does not. Additionally we have found that the quantum phases continue to be dominated by a set of monochromatic regions as it was the case for noninteracting atoms, at least when the induced electric dipolar moments are real.

The paper is organized as follows: Section~\ref{model} derives the model for a system of $N_a$ identical $n$-level atoms interacting with $\ell$ modes of an electromagnetic field, including the atomic dipole-dipole interaction, and particularizes it for $2$- and $3$-level atoms. Section~\ref{variational} constructs the variational energy surface from a complete set of test states which approach the quantum ground state, or any other quantum excited state. Here we focus on the ground state and study its phase diagram. This is applied in section~\ref{2atoms} to the case of $2$-level atoms, and in section~\ref{3atoms} to the case of $3$-level atoms in their different atomic configurations, finding the critical values of the coupling parameters which minimize the energy, and determining the phase diagram in both attractive and repulsive scenarios of the atomic dipole-dipole interaction. In cases where a phase transition exists which defies the Ehrenfest classification, new criteria are proposed, one based on the second Casimir operator and another one based on the maximum Bures distance between neighboring states. Finally, section~\ref{conclusions} summarizes some conclusions. Two appendices present the matter collective operators and the atomic dipole-dipole operator in explicit form.

\section{Model}
\label{model}

\begin{figure}
\begin{center}
\includegraphics[width=0.45\linewidth]{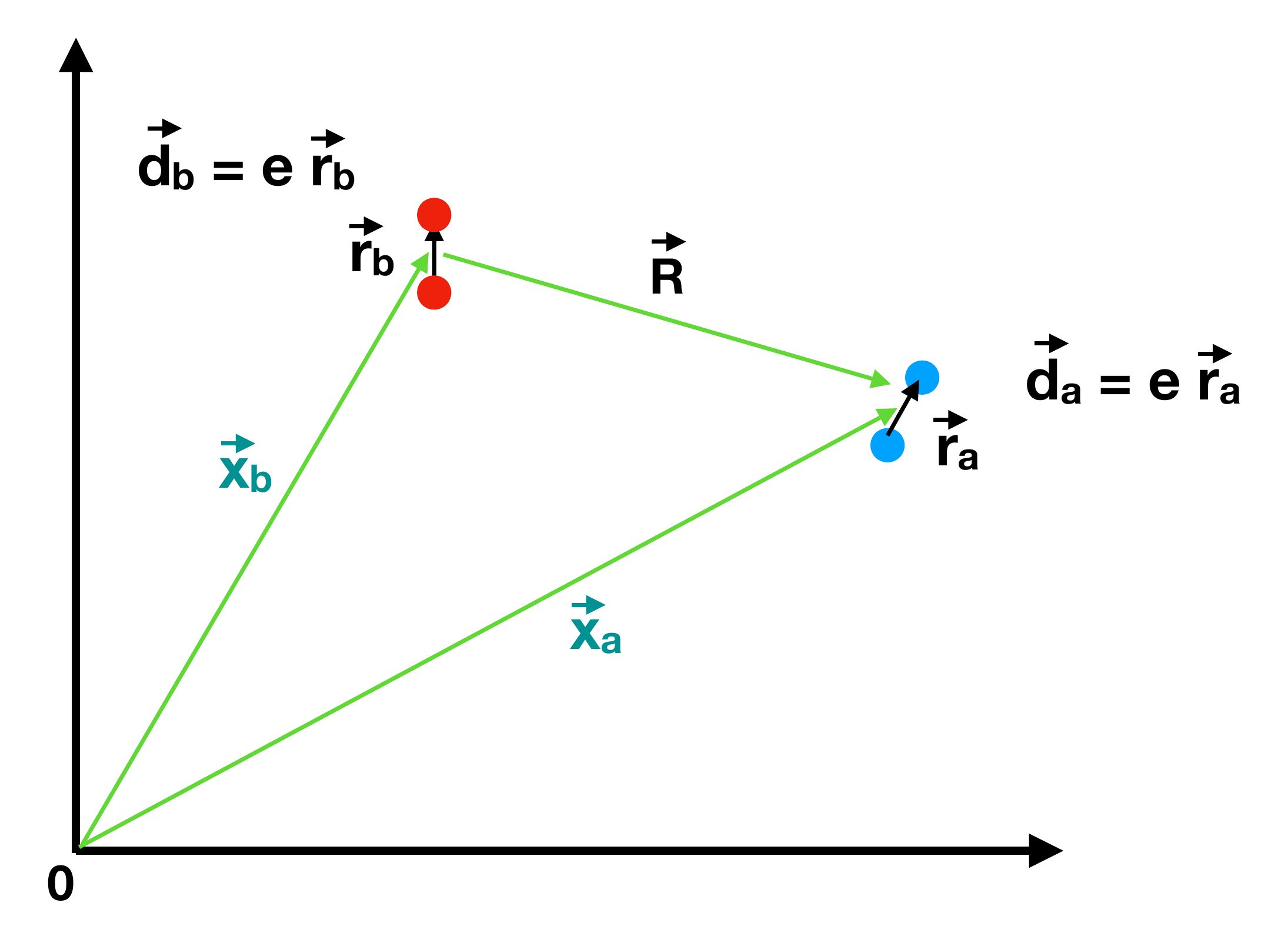} \quad
\includegraphics[width=0.45\linewidth]{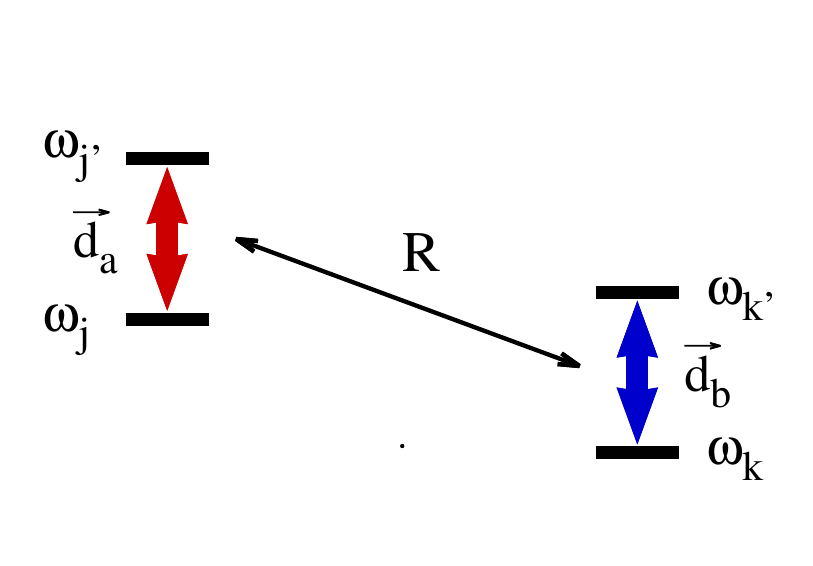}
\caption{(Color online) Left: Schematic of the atomic dipole-dipole interaction. $x_i \ (i=a,b)$ denote the position of the dipoles $\vec{d}_i = e\vec{r_i}$, and $\vec{R}$ their separation. Right: Schematic depiction of the atomic transitions $| \, j, k \rangle \leftrightarrow | \, j^\prime, k^\prime \rangle$ due to the dipole-dipole interaction.}
\label{f.two-dipoles}
\end{center}
\end{figure}

We consider a system of $N_a$ identical $n$-level atoms interacting  with $\ell$ modes of a radiation field, placed in a cavity. The Hamiltonian is composed of three terms 
\begin{equation}\label{eq.H}
\op{H} = \op{H}_D + \op{H}_{mf} + \op{H}_{dd}\,,
\end{equation} 
where $\op{H}_D$ is the diagonal contribution given by ($\hbar=1$) 
\begin{equation}
\op{H}_D = \sum_{s=1}^\ell \Omega_s\ \op{\nu}_s + \sum_{k=1}^n \omega_k \op{A}_{kk}\,;
\end{equation}
the dipolar matter-field interaction is of the form~\cite{cordero13b}
\begin{equation}
\op{H}_{mf} = -\frac{1}{\sqrt{N_a}}\sum_{s=1}^\ell\sum_{j<k}^n \mu_{jk}^{(s)}\left(\op{A}_{jk}+\op{A}_{kj}\left)\right(\op{a}_s^\dag + \op{a}_s\right)\,.
\end{equation}
In these expressions, $\Omega_s$ and $\op{\nu}_s$ are the field frequency and photon number operator, respectively, of mode $s$; $\omega_k$ denotes the energy of the $k$-th atomic level with the convention $\omega_j < \omega_k$ for $j<k$); $\op{a}_s^\dag$ and $\op{a}_s$ are the field creation and annihilation operators; and $\op{A}_{jk}$ is the atomic transition operator between levels $k$ and $j$, which in a bosonic representation $\op{A}_{jk}=\op{b}_j^\dag\op{b}_k$ plays the role of the collective matter operator obeying the unitary algebra in $n$ dimensions, $U(n)$; here, $\op{b}^\dagger_j$ creates an atom in level $j$ and $\op{b}_k$ annihilates one in level $k$ (cf. Eq.~(\ref{eq.comm}). The dipolar matter-field coupling intensity between field mode $s$ and atomic dipole formed by levels $j$ and $k$ is denoted by $\mu_{jk}^{(s)}$.

Atoms do not have permanent dipole moments in their ground state, as the center of charge of the electronic cloud coincides with that of the nucleus. In the presence of an electromagnetic field, however, these centers are displaced and the induced transition dipole moments are responsible for an atomic dipole-dipole interaction.
In second quantization this dipole-dipole interaction takes the form
\begin{equation}\label{wdd1}
\op{H}_{dd} =\frac{1}{2 (N_a-1)} \, \sum_{j,k,j^\prime, k^\prime} \, \langle j, \, k \, | \,\op{W}_{ab} \, | \, j^\prime, k^\prime \rangle \, \op{b}^\dagger_j \, \op{b}^\dagger_k \, \op{b}_{j^\prime}\, \op{b}_{k^\prime}  \, ,
\end{equation}
where bosonic creation $\op{b}_j^\dagger$ and annihilation $\op{b}_k$ operators were used. The factor $1/2$ in eq.~(\ref{wdd1}) compensates the double accounting in the summation, as the particles are indistinguishable. The factor $(N_a-1)$ is included to have an interaction linear in the number of particles. The first index in the bra and the ket states corresponds to the first particle, while the second index corresponds to the second particle.

The set of operators that appear in~(\ref{wdd1}) may be rewritten in terms of the collective matter operators, by means the bosonic commutation relation, as 
\begin{eqnarray}
\op{b}^\dagger_j \, \op{b}^\dagger_k \, \op{b}_{j^\prime}\, \op{b}_{k^\prime} &=& \op{A}_{jj'}\op{A}_{kk'}-\delta_{j'k}\op{A}_{jk'}\nonumber \\[3mm]
&:=& \op{A}_{jj'}\oslash \op{A}_{kk'}\,,
\end{eqnarray}
where we have defined the {\em oslash} product between collective matter operators, which removes the self-interaction terms (see appendix~\ref{ap.collective} for more details).

The dipole-dipole interaction $\op{H}_{dd}$ is obtained from the classical expression~\cite{jackson99} through the standard quantization procedure, which has the form
\begin{equation}\label{eq.W12}
\op{W}_{ab} = \frac{\vec{\op{d}}_a\cdot \vec{\op{d}}_b - 3 (\hat{n}\cdot \vec{\op{d}}_a)(\hat{n}\cdot \vec{\op{d}}_b)}{4\pi \epsilon_0 R^3}\,, 
\end{equation}
where $\vec{\op{d}}_i = e\,\vec{\op{r}}_i,\ (i=a,b)$ are the induced vector operators of the electric dipole moments, $R$ is the separation between the dipoles and $\hat{n}=\vec{R}/R$ (with $\vec{R}=\vec{x}_b-\vec{x}_a$) the unitary vector in the direction from one dipole to another, at positions $\vec{x}_a$ and $\vec{x}_b$ (see Fig.~\ref{f.two-dipoles}). $\epsilon_0$ is the permittivity of vacuum, and for induced magnetic moments $\mu_1,\,\mu_2$, the expression is the same with the replacements $d_i \to \mu_i$, and $\epsilon_0 \to \mu_0$, the magnetic permeability of vacuum (cf., e.g.,~\cite{cohen20}). Without loss of generality, we will consider here that the induced electric dipoles are real.
 
Thus, the two-body matrix elements in Hilbert space are
\begin{equation}\label{eq.gjjpkkp0}
g_{j j'kk'}= \frac{1}{4\, \pi\, \epsilon_0} \left\langle j, \, k \, \left| \, \frac{\vec{\op{d}}_a  \cdot \vec{\op{d}}_b \,  - 3 \,(\hat{n} \cdot \vec{\op{d}}_a ) (\hat{n} \cdot \vec{\op{d}}_b )}{ R^3} \, \right| \, j', \, k' \right\rangle \, ,
\end{equation}
The indices of the dipole-dipole coefficient $g_{j j'kk'}$ refer to the two dipoles  involved in the bra-ket~(\ref{eq.gjjpkkp0}).

For indistinguishable particles, and identifying the expansion components  $\vec{d}_{jk}=\bra j|\vec{\op{d}}_{a,b}|k\ket$ of the dipolar operator in terms of the collective matter operators, viz. the dipolar operator given by  $\vec{\op{d}}=\sum_{j\neq k}^n \vec{d}_{jk}\op{A}_{jk}$, the matrix element in Eq.~(\ref{eq.gjjpkkp0}) reads
\begin{equation}\label{eq.gjjpkkp}
g_{jj'kk'} = \frac{\vec{d}_{jj'}\cdot \vec{d}_{kk'} - 3 (\hat{n}\cdot \vec{d}_{jj'})(\hat{n}\cdot \vec{d}_{kk'})}{4\pi \epsilon_0 R^3}\,,
\end{equation}
where $R$ stands for the average distance between pairs of atoms. The hermiticity of~Eq.(\ref{wdd1}) follows from the relations
\begin{equation}\label{eq.A.grules}
g_{jklm}=g_{lmjk}\,,\qquad g_{jklm}=g_{kjml}^*\,.
\end{equation}
Also, for real dipolar vectors $\vec{d}_{jk}=\vec{d}_{kj}$, one has $g_{jklm}=g_{jkml}=g_{kjlm}$. 

Finally, using the {\it oslash} operator introduced above, we may write the dipole-dipole interaction in a simplified form as
\begin{widetext}
\begin{equation}\label{eq.opWa00}
\op{H}_{dd}= \frac{1}{2(N_a-1)} \sum_{j\neq k}^n\sum_{l \neq m}^n g_{jklm}\op{A}_{jk}\oslash\op{A}_{lm}\,.
\end{equation}
Inserting the different contributions into~(\ref{eq.opWa00}), one may write
the atomic dipole-dipole term in the Hamiltonian as (see appendix~\ref{ap.dd}) 
\begin{eqnarray}\label{eq.Hdd}
\op{H}_{dd} =  \frac{1}{2!}\sum_{j\neq k}^n \op{W}_{jk}^{2-{\rm levels}} + \frac{1}{2!}\sum_{j\neq k\neq l}^n \op{W}_{jlk}^{3-{\rm levels}}
 +\frac{1}{4!} \sum_{j\neq k\neq l \neq m}^n  \op{W}_{jklm}^{4-{\rm levels}}\,,
\end{eqnarray}
\end{widetext}
where the operator $\op{W}_{jk}^{2-{\rm levels}}$ stands for the dipole-dipole contribution of the pair $\vec{d}_{jk} \rightleftharpoons \vec{d}_{jk}$; the operator $\op{W}_{jlk}^{3-{\rm levels}}$ for that of the pair of dipoles $\vec{d}_{jl} \rightleftharpoons \vec{d}_{lk}$ (here the atomic level $\omega_l$ plays the role of an intermediate level, so a prohibited dipolar transition $\vec{d}_{jk}=0$ is possible via the permitted dipolar transitions $\vec{d}_{jl}\neq0$ and $\vec{d}_{lk}\neq0$); and the operator $\op{W}_{jklm}^{4-{\rm levels}}$ corresponds to the contribution of isolated dipoles $\vec{d}_{jk} \rightleftharpoons \vec{d}_{lm}$ (which do not share an energy level). The upper index denotes the number of different atomic levels which contribute to the interaction; hence, the terms $\op{W}_{jlk}^{3-{\rm levels}}$ and $\op{W}_{jklm}^{4-{\rm levels}}$ are zero for $n$-level atoms with $n=2$ and $n\leq3$, respectively.  The set of transitions included in each interaction term is given in table~\ref{t.1}, and shown schematically in figure~\ref{f.t1}. These terms are given in appendix~\ref{ap.dd}. Also, the factors $1/2!$ and $1/4!$ in expression~(\ref{eq.Hdd}) eliminate the double summation due to index reordering. 

\begin{table*}
\caption{Contribution of the atomic transitions to the terms $\op{W}_{jk}^{\rm 2-levels}, \, \op{W}_{jkl}^{\rm 3-levels}$  and $\op{W}_{jklm}^{\rm 4-levels}$ of the atomic dipole-dipole interaction. See also the accompanying figure~\ref{f.t1}.}
\label{t.1}
\vspace{0.2in}
\begin{tabular}{c|c|c| c} \hline
\phantom{aa}Atomic levels \phantom{aa}& \phantom{aa} Interaction \phantom{aa}& \multicolumn{2}{c}{\phantom{aaaa}Atomic Transitions \phantom{aaaa}} \\ \hline
\multirow{2}{\LL}{$\omega_j\,,\omega_k$} & \multirow{2}{\LL}{$\op{W}_{jk}^{\rm 2-levels}$} &
 $j \rightrightarrows k$  &  $j \leftleftarrows k$ \\ \cline{3-4}  & & $j \rightarrow k \rightarrow j$  &  $k \leftarrow j \leftarrow k$ \\ \hline
\multirow{2}{\LL}{$\omega_j\,,\omega_k\,,\omega_l$} & \multirow{2}{\LL}{$\op{W}_{jkl}^{\rm 3-levels}$} &
 $j \leftarrow k \rightarrow l$  &  $j \rightarrow k \leftarrow l$ 
\\ \cline{3-4} 
& & $j \leftarrow k \leftarrow l$  &  $j \rightarrow k \rightarrow l$ 
\\ \hline
\multirow{3}{\LL}{$\omega_j\,,\omega_k\,;\omega_l\,,\omega_m$} & \multirow{3}{\LL}{$\op{W}_{jklm}^{\rm 4-levels}$} &
 $j \rightleftarrows k;\,l \rightleftarrows m$  &    $j \rightleftarrows k;\,m \rightleftarrows l$
\\ \cline{3-4} 
& &  $j \rightleftarrows l;\,k \rightleftarrows m$  &  $j \rightleftarrows l;\,m \rightleftarrows k$ 
\\ \cline{3-4} 
& &  \phantom{aaaa} $j \rightleftarrows m;\,k \rightleftarrows l$\phantom{aaaa}  &  \phantom{aaaa} $j \rightleftarrows m;\,l \rightleftarrows k$\phantom{aaaa} 
\\ \hline
\end{tabular}
\end{table*}
%
\begin{figure*}
\begin{center}
\includegraphics[width=0.9\linewidth]{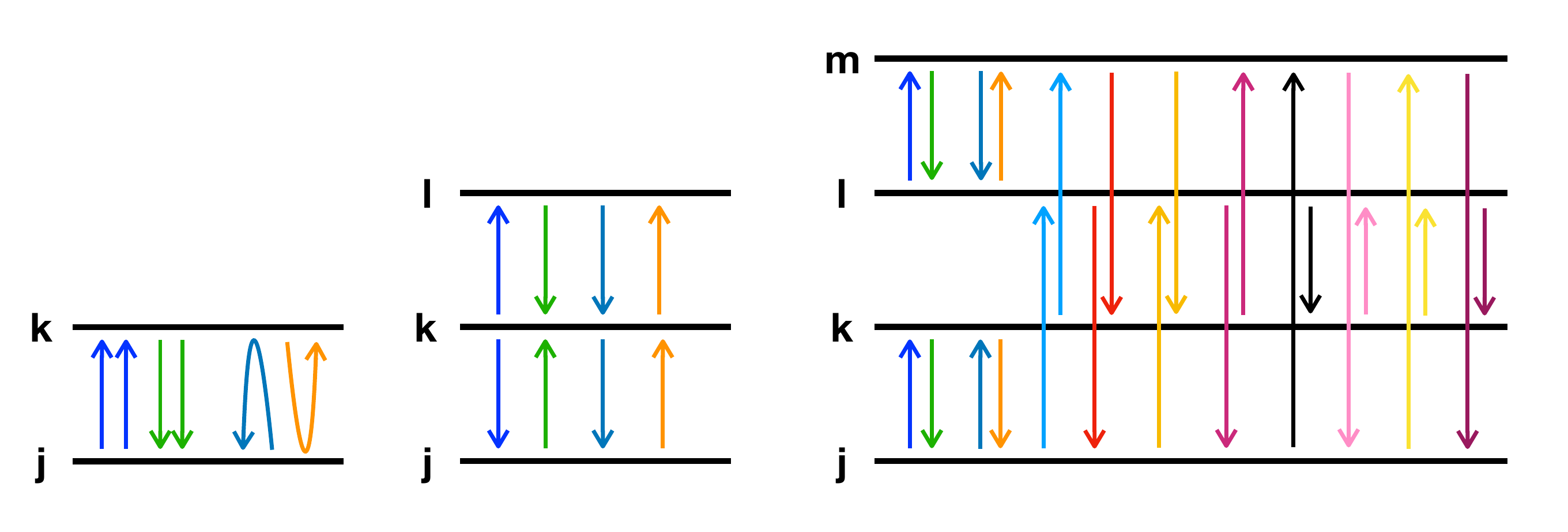}
\caption{(Color online) The set of transitions involved in each term, $\op{W}_{jk}^{\rm 2-levels}, \, \op{W}_{jkl}^{\rm 3-levels}$  and $\op{W}_{jklm}^{\rm 4-levels}$, of the dipole-dipole operator are shown schematically. Each transition is indicated by arrows of the same color. So for two-level atoms (diagram on the left), the blue lines denote the transition $j \rightrightarrows k$, the green lines the transition $j \leftleftarrows k$, the indigo line the transition $j \rightarrow k \rightarrow j$, and the orange line the transition $k \leftarrow j \leftarrow k$.
They are also cumulative; thus, for instance, for the transitions in $\op{W}_{jkl}^{\rm 3-levels}$ we have those shown in the diagram on the left {\it plus} those in the diagram in the middle; similarly, $\op{W}_{jklm}^{\rm 4-levels}$ contains all the transitions in the three diagrams.}
\label{f.t1}
\end{center}
\end{figure*}

Amongst the parameters in the Hamiltonian, we are free to choose $\omega_1=0$ and $\omega_n=1$, i.e., the energies are normalized to the highest atomic level. We also consider systems where only one field mode promotes the transition between a given pair of atomic levels; this constriction is imposed by the condition~\cite{cordero16}
\begin{equation}\label{eq.cond1}
\textrm{if}\quad \mu_{jk}^{(s)}\neq 0 \quad \textrm{then}  \quad\mu_{jk}^{(s')}= 0 \quad\textrm{for all}\quad s'\neq s\,;
\end{equation}
Since the interaction~(\ref{eq.Hdd}) involves the dipole-dipole contribution $g_{jklm}$, we have $g_{jklm}\neq0$ only when $\mu_{jk}^{(s)}\neq0$ and $\mu_{lm}^{(s')}\neq0$ for at least one of the modes $\Omega_s$ and $\Omega_{s'}$.
The first and second order Casimir operators [Eqs.~(\ref{eq.na} and \ref{casimir2})], whose eigenvalues are functions of the number of atoms $N_a$ and the number of levels $n$, are of course constants of motion.

As an example, we now write explicitly the contribution of $\op{H}_{dd}$ for two- and three-level atoms:
\subsubsection*{$2$-level atoms}

For a system of two-level atoms, the Hamiltonian~(\ref{eq.H}) reads
\begin{equation}
\op{H}= \op{H}_D + \op{H}_{mf} + \op{W}_{jk}^{\rm 2-levels}\,,
\end{equation}
where we fix $j<k$ for the atomic levels $\omega_j < \omega_k$ respectively.

\subsubsection*{$3$-level atoms}

$3$-level atoms present three different atomic configurations ($\Xi$, $\Lambda$ and $V$), according to which atomic transitions are prohibited.
\begin{itemize}
\item
For the $\Xi$-configuration, the dipolar transition $\vec{d}_{13}=0$ is prohibited, and the Hamiltonian takes the form
\begin{eqnarray}
\op{H}_\Xi&=& \op{H}_D + \op{H}_{mf}\nonumber\\[2mm]
&+& \op{W}_{12}^{2-{\rm levels}} +  \op{W}_{23}^{2-{\rm levels}}  + \op{W}_{123}^{3-{\rm levels}} \,.
\end{eqnarray}
The intermediate atomic level $\omega_2$ may promote the transition $\omega_1\rightleftharpoons \omega_3$. The set of nonzero dipolar-dipolar strengths is $\{g_{1212},g_{1221},g_{2323},g_{2332},g_{1232},g_{1223}\}$ together with their complex conjugates, obtained as $g_{jklm}=g_{kjml}^*$, cf. Eq.~(\ref{eq.A.grules}).

\item
For the $\Lambda$-configuration it is the dipolar transition $\vec{d}_{12}=0$ which is prohibited, and the Hamiltonian takes the form
\begin{eqnarray}
\op{H}_\Lambda&=& \op{H}_D + \op{H}_{mf}\nonumber\\[2mm]
&+& \op{W}_{13}^{2-{\rm levels}} +  \op{W}_{23}^{2-{\rm levels}}  + \op{W}_{132}^{3-{\rm levels}} \,.
\end{eqnarray}
The atomic level $\omega_3$ serves as an intermediate level which may promote the transition $\omega_1\rightleftharpoons \omega_2$. The set of nonzero dipolar-dipolar strengths is $\{g_{1313},g_{1331},g_{2323},g_{2332},g_{1323},g_{1332}\}$ together with their complex conjugates obtained by Eq.~(\ref{eq.A.grules}). 

\item
For the $V$-configuration the dipolar transition $\vec{d}_{23}=0$ is prohibited, and the Hamiltonian is
\begin{eqnarray}
\op{H}_V&=& \op{H}_D + \op{H}_{mf}\nonumber\\[2mm]
&+& \op{W}_{12}^{2-{\rm levels}} +  \op{W}_{13}^{2-{\rm levels}}  + \op{W}_{213}^{3-{\rm levels}} \,.
\end{eqnarray}
The atomic level $\omega_1$ acts here as an intermediate and may promote the transition $\omega_2\rightleftharpoons \omega_3$. The set of nonzero dipolar-dipolar strengths is $\{g_{1212},g_{1221},g_{1313},g_{1331},g_{2131},g_{2113}\}$ together with their complex conjugates obtained by Eq.~(\ref{eq.A.grules}). 

\end{itemize}

In a recent work~\cite{civitarese10} the case where equal contributions of the form  $g_{jppj}=g_{jppk}=g$ for all $j,\,k$, was considered (other terms were neglected). It was shown that the dipole–-dipole interactions act against the appearance of atomic squeezing, and also that an increase in the mean value of the number of photons of the initial state smears out the effect.

\section{Variational energy surface}
\label{variational}

The variational solution involves a test state which approaches the quantum ground or desired excited state, and which depends on a set of parameters $z_i$. The corresponding energy surface is obtained by taking the expectation value of the Hamiltonian and minimizing with respect to the parameters $z_i$ of the test state. In this work we focus on the ground state and take as test state the direct product of coherent states for both the matter and the field contributions. Clearly, this test state presents no matter-field entanglement, but it yields a good description of the minimum energy surface, as well as some expectation values of the physical quantities, and the phase diagram together with the order of the phase transitions. 

\subsubsection*{Coherent matter state}

The coherent matter state is defined as~\cite{iachello87}
\begin{equation}\label{eq.sgamma}
|\vec{\gamma}\ket =\frac{1}{\sqrt{N_a!}}\left[\op{\Gamma}^\dag\right]^{N_a} |0\ket_m\,,
\end{equation}
where $\vec{\gamma}=(\gamma_1,\dots,\gamma_n)$ and  $||\vec{\gamma}||:=(|\gamma_1|^2+|\gamma_2|^2+\cdots+|\gamma_n|^2)^{1/2}$.
The operator $\op{\Gamma}^\dag$ is 
\begin{equation}\label{eq.ogamma}
\op{\Gamma}^\dag=\frac{\gamma_1\op{b}_1^\dag+\gamma_2\op{b}_2^\dag+\cdots+\gamma_n\op{b}_n^\dag}{||\vec{\gamma}||}\,,
\end{equation}
and, using the bosonic realization  $[\op{b}_j,\op{b}_k^\dag]=\delta_{jk}$ it is immediate that the relationship $[\op{\Gamma},\op{\Gamma}^\dag]=1$ is fulfilled; hence, the state (\ref{eq.sgamma}) is normalised.  It is straightforward to show that 
\begin{equation}
\left[\op{b}_k,\op{\Gamma}^\dag\right]= \frac{\gamma_k}{||\vec{\gamma}||}\,,
\end{equation}
which for any number of atoms generalizes to
\begin{equation}
\left[\op{b}_k,(\op{\Gamma}^\dag)^{N_a}\right]= N_a\frac{\gamma_k}{||\vec{\gamma}||} (\op{\Gamma}^\dag)^{N_a-1}\,.
\end{equation}
The relations above are useful in order to find the matrix elements of the collective matter operators. The linear contribution is 
\begin{equation}
\bra\vec{\gamma}|\op{A}_{jk}|\vec{\gamma}\ket = N_a \frac{\gamma_j^*\gamma_k}{||\vec{\gamma}||^2}\,,
\end{equation}
and the quadratic contribution 
\begin{eqnarray}
\bra\vec{\gamma}|\op{A}_{jk}\op{A}_{lm}|\vec{\gamma}\ket =&& N_a(N_a-1) \frac{\gamma_j^*\gamma_k \gamma_l^*\gamma_m}{||\vec{\gamma}||^4} \nonumber\\
&& + \delta_{kl} N_a \frac{\gamma_j^*\gamma_m}{||\vec{\gamma}||^2}\,,
\end{eqnarray}
where the last term corresponds to the self-interactions, and vanishes in the dipole-dipole interaction.
\vspace{0.15in}

\subsubsection*{Coherent field state}

The coherent field state for $\ell$ modes is given by the direct product of coherent states for each mode, as follows~\cite{klauder85, ali14},
\begin{equation}
|\vec{\alpha}\ket := |\alpha_1\ket\otimes|\alpha_1\ket\otimes\cdots\otimes|\alpha_\ell\ket\,,
\end{equation}
where $\vec{\alpha}=\{\alpha_1,\dots,\alpha_\ell\}$. For each mode $s=1,\dots,\ell$ the coherent state satisfies  $\op{a}_s |\alpha_s\ket = \alpha_s |\alpha_s\ket$, and hence
\begin{equation}
\bra \vec{\alpha}|\op{a}_s |\vec{\alpha}\ket = \alpha_s \,, \qquad \bra \vec{\alpha}|\op{a}_s^\dag |\vec{\alpha}\ket = \alpha_s^* \,,
\end{equation}
while the expectation value of the number operator for each mode is
\begin{equation}
\bra \vec{\alpha}|\op{a}_s^\dag\op{a}_s |\vec{\alpha}\ket =\bra \vec{\alpha}|\op{\nu}_s |\vec{\alpha}\ket= |\alpha_s|^2\,.
\end{equation}
\vspace{0.1in}

From the expressions above, and writing for the complete test state the direct product of the coherent states for field and matter, 
\begin{equation}
|\vec{\alpha},\vec{\gamma}\ket := |\vec{\alpha}\ket\otimes|\vec{\gamma}\ket\,,
\end{equation}
the variational energy surface per atom ${\cal E}:= \bra\vec{\alpha},\vec{\gamma}| \op{H}|\vec{\alpha},\vec{\gamma}\ket/N_a$, as a function of $\alpha_s =R_s e^{i\theta_s}$, $\gamma_k=\varrho_k e^{i \phi_k}$, and parameters of the Hamiltonian, reads
\begin{eqnarray}\label{eq.Ev}
{\cal E}&=& \frac{1}{N_a}\sum_{s=1}^\ell \Omega_s\ R_s^2 + \sum_{k=1}^n \omega_k \frac{\varrho_k^2}{||\vec{\gamma}||^2}\nonumber \\[3mm]
&-&  \frac{4}{\sqrt{N_a} }\sum_{s=1}^\ell\sum_{j<k}^n \mu_{jk}^{(s)}\frac{\varrho_j\varrho_k R_s}{||\vec{\gamma}||^2} \cos(\phi_{jk})\cos(\theta_{s})\nonumber\\[3mm]
&+& \frac{1}{N_a} \bra\vec{\alpha},\vec{\gamma}| \op{H}_{dd} |\vec{\alpha},\vec{\gamma}\ket \,,\quad
\end{eqnarray}
where $\phi_{jl}=\phi_l-\phi_j$. The last term in~(\ref{eq.Ev}) corresponds to the atomic dipole-dipole interaction per particle  ${\cal E}_{dd}$, and has the form 
\begin{widetext}
\begin{eqnarray}
{\cal E}_{dd} &=& \frac{1}{||\vec{\gamma}||^4} \sum_{j<k} {\rm Re}\, [g_{jkjk}\,  e^{2i \phi_{jk}} +  \,g_{jkkj} ]\, \varrho_j^2\varrho_k^2+ \frac{2}{||\vec{\gamma}||^4} \sum_{j<k; j\neq p\neq k} {\rm Re}\,[ g_{jpkp} \, e^{i(\phi_{jp}+\phi_{kp})}+g_{jppk}\,e^{i\phi_{jk}} ]\,\varrho_j \varrho_k \varrho_p^2
\nonumber \\[3mm]
&+& \frac{2}{||\vec{\gamma}||^4}\sum_{j<k<l<m} {\rm Re}\,[
g_{j k l m} \,e^{i(\phi_{jk}+\phi_{lm})} +
g_{j k m l} \,e^{i(\phi_{jk}-\phi_{lm})}   +
g_{j l k m} \,e^{i(\phi_{jl}+\phi_{km})}+
g_{j l m k} \,e^{i(\phi_{jl}-\phi_{km})}\nonumber \\ && +
g_{j m k l} \,e^{i(\phi_{jm}+\phi_{kl})}+
g_{j m l k} \,e^{i(\phi_{jm}-\phi_{kl})} ] \, \varrho_j \varrho_k \varrho_l \varrho_m \,.
\label{eq-dip-dip}
\end{eqnarray}
\end{widetext}
Here, we used the fact that $g_{jklm}=g_{lmjk}$ and $ g_{jklm}=g_{kjml}^*$ in order to simplify the expression.

By simple inspection, one may note that the energy surface has minima at the critical values $\theta_s^c=0,\pi$ and $R_s^c = \sqrt{N_a}\, r_s^c$, with 
\begin{equation}\label{eq.rsc1}
r_s^c =  2\sum_{j<k}^n \frac{\mu_{jk}^{(s)}}{\Omega_s}\frac{\varrho_j^c\varrho_k^c }{||\vec{\gamma}^c||^2} \cos(\phi_{jk}^c)\cos(\theta_{s}^c)\,, \quad r_s^c \geq 0\,,
\end{equation}
and $\phi_{jl}^c=\phi_l^c-\phi_j^c$.

In a similar fashion, for the fixed values $\varrho_1=1$ and $\phi_1=0$, and supposing, without loss of generality, real values for the dipolar ($\mu$) and dipole-dipole ($g_{jklm}$) strengths, one finds the critical values for the phase $\phi_j$ to be  $\phi_j^c=0,\,\pi$.

After substitution of the critical values $\theta_s^c, \,\phi_j^c$, and fixing $\varrho_1=1$ and $\phi_1=0$, we obtain a family of energy surfaces $E(\varrho;\theta^c,\phi^c)$ for $\varrho=(\varrho_2\,\dots\,\varrho_n),\, \theta^c=(\theta^c_1\,\dots\,\theta^c_\ell)$ and $\phi^c=(\phi^c_2\,\dots\,\phi^c_n)$; appropriate values for $\theta_s^c$ and $\phi_j^c$ should be selected in order to satisfy $r_s^c\geq0$ in Eq.~(\ref{eq.rsc1}). The minimum energy surface is then obtained by calculating the critical points $\varrho_j^c$, which is done numerically in general.

\section{Two-level atoms}
\label{2atoms}

For two-level atoms the expression of the energy surface reads
\begin{eqnarray}
{\cal E}&=& \Omega_s\,r_s^2  + \frac{\omega_j \, \varrho_j^2 + \omega_k \,\varrho_k^2}{\varrho_j^2 +\varrho_k^2}  - 4 \frac{\mu_{jk}\,r_s\varrho_j\varrho_k\cos(\theta_s)\cos(\phi_{jk})}{\varrho_j^2 +\varrho_k^2}\nonumber \\[3mm]
&+&   [g_{jkjk}\,  \cos(2\, \phi_{jk}) +  \,g_{jkkj} ]\, \frac{\varrho_j^2\varrho_k^2}{(\varrho_j^2+\varrho_k^2)^2}\,,
\end{eqnarray}
with $j<k$. The critical values of the corresponding energy surface Eq.~(\ref{eq.Ev}) must satisfy 
\begin{equation}
\mu_{jk} \cos(\phi_{jk}^c)\cos(\theta_s^c) = |\mu_{jk}|\,,
\end{equation}
and, fixing $\phi_j=0$ and $\varrho_j=1$, one finds two solutions
\begin{eqnarray}\label{eq.rhokc}
\varrho_k^c = 0 ,\quad \textrm{and}\quad  \varrho_k^c =  \sqrt{\frac{x_{jk}^2 -y_{jk}}{x_{jk}^2 -y_{jk}+2}}\,.
\end{eqnarray}
Here, we have used a dimensionless matter-field coupling intensity $x_{jk}$, defined as
\begin{equation}
x_{jk}=\frac{\mu_{jk}}{\mu_{jk}^c}\,,\qquad \mu_{jk}^c = \frac{1}{2}\sqrt{\Omega_s \,\omega_{jk}}\,,
\end{equation}
with $\mu_{jk}^c$ the critical value of the coupling constant when the atomic dipole-dipole interaction is neglected, and where $\omega_{jk}=|\omega_k-\omega_j|$; we have also defined
\begin{equation}
y_{jk}=\frac{\omega_{jk}+g}{\omega_{jk}}\,, \quad g=g_{jkkj}+g_{jkjk}\,,
\label{xcrit}
\end{equation}
to simplify the notation.

\begin{figure}
\begin{center}
\includegraphics[width=0.80\linewidth]{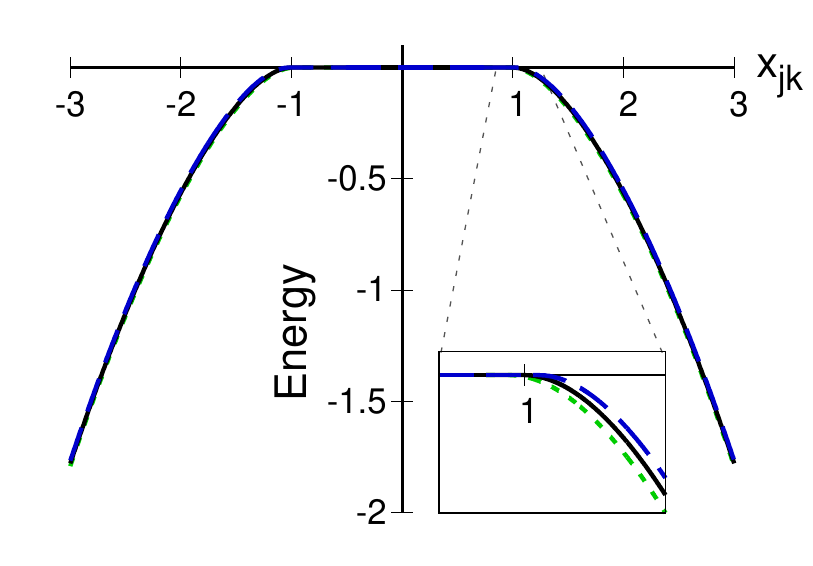}
\caption{(colour online) Minimum energy as a function of the matter-field coupling $x_{jk}$, for two-level atoms interacting with a single mode of an electromagnetic field. The solid line corresponds to the case without dipole-dipole interaction $g=0$; the repulsive case $g=0.05$ (dashed line) and the attractive case $g=-0.05$ (dotted line). Inset shows a zoom around the value $x_{jk}=1$ where the transition appears. The parameters are $\omega_j=0\,,\omega_k=1$ for the atomic levels, and $\Omega=1$ for the field frequency.}\label{f.emin2la}
\end{center}
\end{figure}

For values $x_{jk}^2-y_{jk}\leq 0$ one finds only one critical value $\varrho_k=0$, for which the energy surface has the constant value $\omega_j$. When $x_{jk}^2-y_{jk} > 0$ we have two critical values~(\ref{eq.rhokc}), in this case the energy surface has a dependence on the matter-field dipolar strength $x_{jk}$. After minimizing one finds
\begin{equation}
E_{\rm min} = \left\{ \begin{array}{l l} \omega_j ;& x _{jk}^2 <y_{jk} \\[3mm]
\omega_j - \displaystyle{\frac{[x_{jk}^2 - y_{jk}]^2}{4(x_{jk}^2 - y_{jk}+1) }} \omega_{jk};& x_{jk}^2 \geq y_{jk} 
\end{array}\right.\,,
\end{equation}
This relationship, $E_{\rm min}$ vs. $x_{jk}$, is shown in figure~\ref{f.emin2la}. The solid line (black) corresponds to the case $g=0$ without atomic dipole-dipole interaction; repulsive $g=0.05$ (dashed line, blue) and attractive $g=-0.05$ (dotted line, green) cases are also shown. Due to the minuteness of this interaction when compared with the dipolar matter-field interaction, the difference for dissimilar values of $g$ is difficult to appreciate. We have zoomed around the value $x_{jk}=1$ (see figure inset) where the transition into the collective region appears, in order to make this difference clear. In what follows, we will consider unnaturally large values for the atomic dipole-dipole coupling parameter $g$ so that its effect may be appreciated; when studying actual realistic systems these values (and their effects) must be scaled down accordingly.

%

\begin{figure*}
\begin{center}
\includegraphics[width=0.48\linewidth]{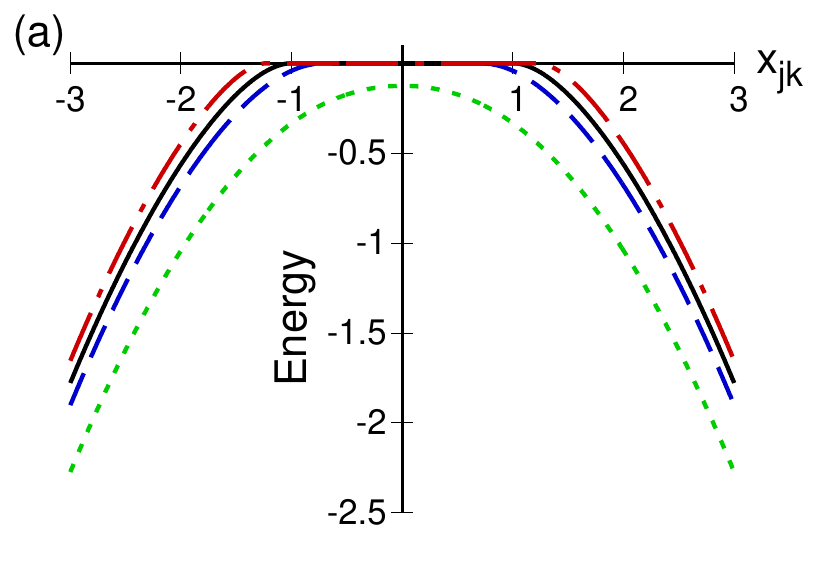}
\includegraphics[width=0.48\linewidth]{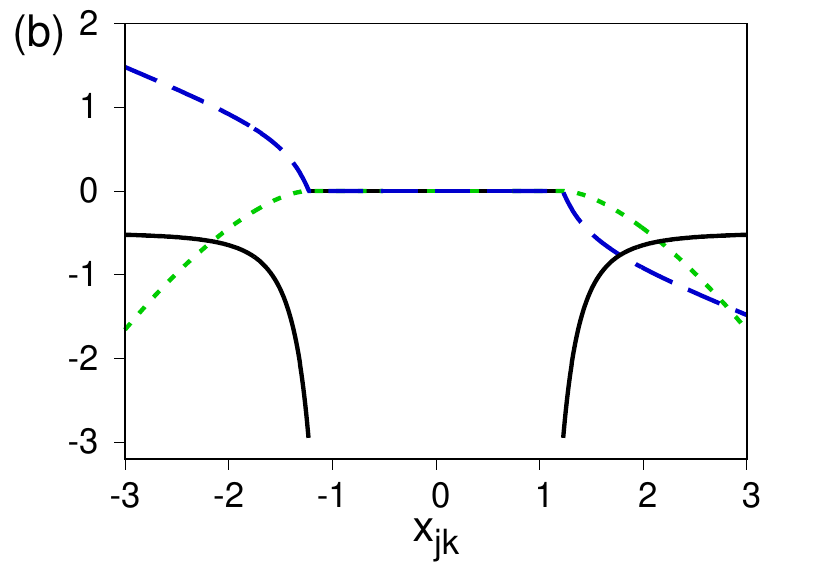}
\caption{(colour online) (a) Minimum energy as a function of the matter-field coupling $x_{jk}$, for two-level atoms interacting with a single mode of an electromagnetic field. The solid line corresponds to the case $g=0$ without dipole-dipole interaction; the repulsive case $g=0.5$ (dash-dot line), and two attractive cases $g=-0.5$ (dashed line) and $g=-2$ (dotted line) are also shown (the latter in a regime of very strong attractive interaction).  (b) Minimum energy (dotted line) and its first (dashed line) and second  (solid line) derivatives. The parameters are $\omega_j=0\,,\omega_k=1$ for the atomic levels, and $\Omega=1$ for the field frequency.}\label{f.emin2l}
\end{center}
\end{figure*}

The minimum energy for different (larger) values of the dipolar coupling strength is plotted in figure~\ref{f.emin2l}. For values of $g$ such that $y_{jk}>0$, the critical points $x_{jk}^c=\pm \sqrt{y_{jk}}$ divide the {\em normal region} $x_{jk}^2<(x_{jk}^c)^2$ from the {\em collective region} $x_{jk}^2>(x_{jk}^c)^2$.  One should note that for the case without dipole-dipole interaction, $g=0$, (solid line in figure~\ref{f.emin2l}(a)) the critical points occur at $(x_{jk}^c)^2=1$, while in the attractive case $g<0$ (dashed line in figure~\ref{f.emin2l}(a)) one has $(x_{jk}^c)^2<1$, i.e., the normal region decreases. Correspondingly, for the repulsive case $g>0$ (dot-dash line in figure~\ref{f.emin2l}(a)) the normal region increases, as we have $(x_{jk}^c)^2>1$. The anomalous behaviour is the {\em strong attractive regime}, this is characterised by values of $g$ such that $y_{jk}\leq0$,  when the normal region vanishes completely (dotted line in Fig.~\ref{f.emin2l}(a)). It is important to note that, for large matter-field coupling $x_{jk}^2\gg y_{jk}$, the minimum energy surface $E_{\rm min}$ tends to that without the atomic dipole-dipole interaction; in other words, the effect of the dipole-dipole terms on the energy surface is seen mainly in a vicinity of the normal region.

The order of the transition may be determined using the Ehrenfest classification~\cite{gilmore93}, which involves the derivatives of the energy surface. We exemplify the case $g=0.5$ in figure~\ref{f.emin2l}(b), showing, respectively, the first (dashed-line) and second derivatives (solid line) of the energy. Since the second derivative presents a discontinuity at the critical point $x_{jk}^c$, a second order transition occurs at that location.

\section{Three-level atoms}
\label{3atoms}

For three-level atomic systems interacting dipolarly with a two-mode electromagnetic field in a cavity, the atomic dipole-dipole interaction can be obtained from expression~(\ref{eq.opWa00}) or~(\ref{B2}). For the case of real induced dipole moments one has only to consider the real coupling strengths $g_{1212},\, g_{1313},\, g_{2323}$ for two-level interactions, and $g_{1213},\, g_{1232},\, g_{1323}$ for those associated to three-level interactions. Thus the induced dipole-dipole interaction for three-level atoms takes the form,
\begin{widetext}
\begin{eqnarray}
\bm{H}_{dd} &=& \frac{g_{1212}}{2 \, (N_a-1)}   \left\{(\bm{A}_{12}+\bm{A}_{21})^2 -\bm{A}_{11} -\bm{A}_{22} \right\} 
 +  \frac{g_{1313}}{2 \, (N_a-1)} \left\{(\bm{A}_{13}+\bm{A}_{31})^2 -\bm{A}_{11} -\bm{A}_{33} \right\}  \nonumber \\
 && +  \frac{g_{2323}}{2 \, (N_a-1)} \left\{ (\bm{A}_{23}+\bm{A}_{32})^2 -\bm{A}_{22} -\bm{A}_{33}  \right\} 
 +  \frac{g_{1213} }{N_a-1} \left\{ \bm{A}_{12}\, \bm{A}_{13} + \bm{A}_{31}\, \bm{A}_{21} +  \bm{A}_{13}\, \bm{A}_{21} +  \bm{A}_{12}\, \bm{A}_{31} \right\} \, \\ 
 && +   \frac{g_{1232} }{N_a-1} \left\{ \bm{A}_{12}\, \bm{A}_{32} + \bm{A}_{23}\, \bm{A}_{21} +  \bm{A}_{23}\, \bm{A}_{12} +  \bm{A}_{21}\, \bm{A}_{32} \right\} \, 
 +   \frac{g_{1323} }{N_a-1} \left\{ \bm{A}_{13}\, \bm{A}_{23} + \bm{A}_{32}\, \bm{A}_{31} +  \bm{A}_{32}\, \bm{A}_{13} +  \bm{A}_{31}\, \bm{A}_{23} \right\} \,.\nonumber 
\label{eq3levels}
\end{eqnarray}
\end{widetext}
Notice that for the different atomic configurations one has at most three real parameters; in the case of the $\Lambda$ configuration, for instance, we have the coupling strengths $g_{1313}, \, g_{2323}$ and $g_{1323}$.

The corresponding variational energy surface for the dipole-dipole interaction may be obtained by taking the expectation value of~(\ref{eq3levels}) with respect the variational state $| \gamma_1, \, \gamma_2, \, \gamma_3\rangle \otimes | \alpha_1,\, \alpha_2\rangle$, or from the general expression~(\ref{eq-dip-dip}) by considering real induced dipole moments together with three-level atomic systems and a two-mode electromagnetic field.  The resulting expressions for the $\Lambda$, $V$ and $\Xi$ atomic configuration are given by
\begin{widetext}
\begin{eqnarray}
{\cal E}^{(\Lambda)}_{dd} &=& \frac{g_{1313} \, \rho^2_3 \, (1 + \cos{2\, \phi_3})}{(1 + \rho^2_2+\rho^2_3)^2} + \frac{g_{2323} \, \rho^2_3 \, \rho^2_2  \, (1 + \cos{2(\phi_3}-\phi_2))}{(1 + \rho^2_2+\rho^2_3)^2} 
 +  \frac{2\, g_{1323} \, \rho^2_3 \, \rho_2 \, (\cos(2\, \phi_3-\phi_2) + \cos{\phi_2})}{(1 + \rho^2_2+\rho^2_3)^2} \, ,  \\[5mm]
{\cal E}^{(V)}_{dd} &=& \frac{g_{1212} \, \rho^2_2 \, (1 + \cos{2\, \phi_2})}{(1 + \rho^2_2+\rho^2_3)^2} + \frac{g_{1313} \, \rho^2_3 \, (1 + \cos{2\, \phi_3})}{(1 + \rho^2_2+\rho^2_3)^2} 
 +  \frac{4\, g_{1213} \, \rho_2 \, \rho_3 \,  \cos\phi_2 \, \cos{\phi_3}}{(1 + \rho^2_2+\rho^2_3)^2} \, , \\[5mm]
{\cal E}^{(\Xi)}_{dd} &=& \frac{g_{1212} \, \rho^2_2 \, (1 + \cos{2\, \phi_2})}{(1 + \rho^2_2+\rho^2_3)^2} + \frac{g_{2323} \, \rho^2_2 \, \rho^2_3 \, (1 + \cos{2\,(\phi_3}-\phi_2))}{(1 + \rho^2_2+\rho^2_3)^2} 
 +  \frac{2\, g_{1232} \, \rho^2_2 \, \rho_3 \, (\cos(2\, \phi_2-\phi_3) + \cos{\phi_3})}{(1 + \rho^2_2+\rho^2_3)^2} \, .
\end{eqnarray}
\end{widetext}

For systems of $3$-level atoms interacting with two modes of electromagnetic field, the critical values of the phases ({\it vide supra}) are $\theta_s^c =0,\, \pi$ and $\phi_{k}^c=0,\, \pi$, for which the relationship $\mu_{jk}^{(s)} \cos(\theta_s^c)\cos(\phi_{jk}^c)>0$ is satisfied, and where we defined $\phi_{jk}^c = \phi_k^c-\phi_j^c$. Also, the critical values $r_s^c$ associated to the field are given as functions of the critical values $\varrho_k^c$ of the matter [cf. Eq.~(\ref{eq.rsc1})].  These values must be calculated numerically, except when the dipole-dipole interaction is neglected, since in this latter case we have an analytical solution~\cite{cordero15}.

In this work we calculate the critical values for the three atomic configurations ($\Xi$, $\Lambda$ and $V$) and obtain the corresponding separatrix; we fix in all cases the double resonant condition, i.e., the field frequencies are given by $\Omega_1=\omega_{jk}$ and $\Omega_2=\omega_{lm}$. The atomic levels satisfy the condition $\omega_1<\omega_2<\omega_3$  with $\omega_1=0$ and $\omega_3=1$. We take $(j,k,l,m)=(1,2,2,3)$ and the value $\omega_2=3/4$ for the $\Xi$-configuration,  $(j,k,l,m)=(1,3,2,3)$ and $\omega_2=1/4$ for the $\Lambda$-configuration, and $(j,k,l,m)=(1,2,1,3)$ and $\omega_2=3/4$ for the $V$-configuration. The values considered for the dipolar-dipolar strength $g_{jklm}$, assuming real dipolar vectors $\vec{d}_{jk}=\vec{d}_{kj}$, are given in table~\ref{t.parg}.
\begin{table}
\caption{Values for the dipole-dipole strength  $g_{\pm s}$ used in the numerical calculation of the minimum energy surface. The indices are $(j,k,l,m)=(1,2,2,3)$ for the $\Xi$-configuration, $(j,k,l,m)=(1,3,2,3)$ for the $\Lambda$-configuration, and $(j,k,l,m)=(1,2,1,3)$ for the $V$-configuration. We have used the relationship $g_{jklm}=g_{jkml}$ assuming real dipolar vectors $\vec{d}_{jk}$. \\} \label{t.parg}
\begin{tabular}{c l l l} \hline
& \phantom{aaaa}$g_{jkjk}$ & \phantom{aaaa}$g_{lmlm}$ & \phantom{aaaa}$g_{jklm}$\\ \hline
$g_{\pm 1}$ & \phantom{aaa}$\pm 0.1$& \phantom{aaa}$\pm 0.04$ & \phantom{aaa}$\pm 14 \,\sqrt{10^{-5}}$\\
$g_{\pm 2}$ & \phantom{aaa}$\pm 0.3 $& \phantom{aaa}$\pm 0.2$ & \phantom{aaa}$\pm  14 \,\sqrt{3/2} \times 10^{-2}$\\
$g_{\pm 3}$ & \phantom{aaa}$\pm 1.0$& \phantom{aaa}$\pm 0.4$ & \phantom{aaa}$\pm 140 \,\sqrt{10^{-5}}$\\ \hline
\end{tabular}
\end{table}
%

\begin{figure*}
\begin{center}
\includegraphics[width=0.48\linewidth]{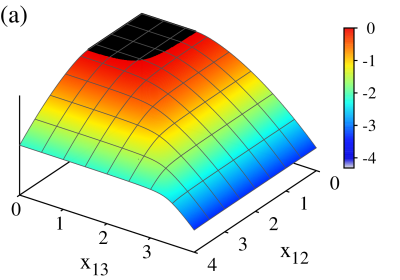}
\includegraphics[width=0.48\linewidth]{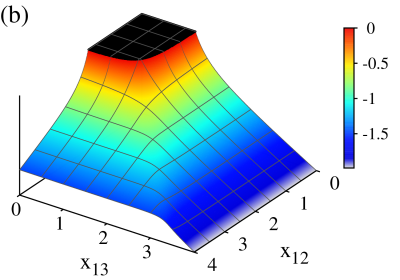}\\
\includegraphics[width=0.48\linewidth]{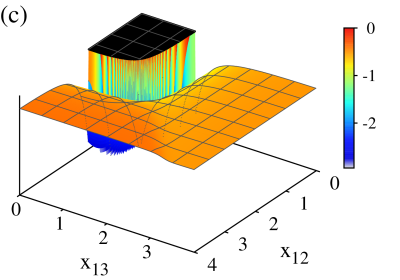}
\includegraphics[width=0.48\linewidth]{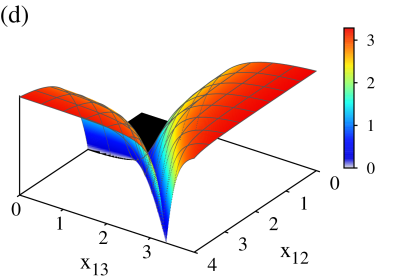}
\caption{(colour online) $V$-configuration with fixed values of $g_{3}$. (a) shows the minimum energy surface, (b) its first derivative [Eq.~(\ref{eq.dE})], (c) its second derivative [Eq.~(\ref{eq.d2E})], and (d) the difference between the second order Casimir operators of the subsystems. Parameters used are discussed in the text.} \label{f.Vexample}
\end{center}
\end{figure*}

In order to exemplify how to obtain the separatrix, we consider explicitly the particular case of the $V$-configuration with a repulsive dipole-dipole strength $g_3$. The set of critical points $\varrho_2^c$ and $\varrho_3^c$ are evaluated numerically and inserted into the expression for the minimum energy; the result is shown in Fig.~\ref{f.Vexample}(a). The normal region, where $E_{min}=0$, is colored in black. The separatrix is found by calculating the first derivatives of the energy surface, as
\begin{equation}\label{eq.dE}
\delta E =: \frac{\partial E}{\partial x_{jk}} + \frac{\partial E}{\partial x_{lm}}\,,
\end{equation}
which is shown in Fig.~\ref{f.Vexample}(b). It is a continuous surface. We calculate the second order derivative as
\begin{equation}\label{eq.d2E}
\delta^2 E =: \frac{\partial \delta E}{\partial x_{jk}} + \frac{\partial \delta E}{\partial x_{lm}}\,,
\end{equation}
which is discontinuous [cf. Fig.~\ref{f.Vexample}(c)]. The loci form a separatrix which splits the normal from the collective region; in fact, this discontinuity shows that a second order transition occurs at these points for the $V$-configuration.

In Fig.~\ref{f.Vexample}(c), the slight undulation (observed by a small change in the orange hue of the surface) within the collective region in the second derivative of the minimum energy surface, is a signature of a kind of transition due to a change of subspaces formed by $2$-level atoms, as was discussed recently for the case without dipole-dipole interaction $g=0$~\cite{cordero15}, from one subspace in which one of the radiation modes dominates to another subspace where the other mode dominates. This change grows as $g\to0$, and gives a discontinuity when $g=0$. However, for values $g\neq0$ the second derivative remains continuous, as well as derivatives of higher order; in other words, the Ehrenfest classification does not provide a criterion to determine that the transition exists. In this work, we propose to consider the second order Casimir operator corresponding to each $2$-level subsystem in order to label this transition ({\it vide infra}).

The second order Casimir operator for a system of $N_a$ particles of $n$-levels is given by
\begin{equation}\label{eq.casimir}
\sum_{j,k=1}^n \op{A}_{kj}\op{A}_{jk} = N_a\,(N_a+n-1)\,.
\end{equation}
In particular, when only two levels are considered, we may define 
\begin{equation}\label{eq.cjk}
C_{jk}=:\op{A}_{jj}\op{A}_{jj}+\op{A}_{jk}\op{A}_{kj}  +\op{A}_{kj}\op{A}_{jk}  +\op{A}_{kk}\op{A}_{kk} \,,
\end{equation}
which coincides with the second order Casimir operator for $2$-levels. Therefore, the expectation value $\bra \psi | C_{jk} | \psi\ket $ will be close to $N_a(N_a+1)$ when the bulk of the contribution to the state $|\psi\ket$ is given by the basis of the sub-system of the two levels $(j,k)$.  Since the variational solution is independent of $N_a$, we fix for this calculation $N_a=2$ and consider the absolute value of the difference of the second order Casimir operator of each subsystem
\begin{equation}\label{eq.dC}
\delta C =:  |\bra \psi |C_{jk} - C_{lm}|\psi\ket |\,,
\end{equation}
where $|\psi\ket$ stands for the ground state.

This quantity is plotted in figure~\ref{f.Vexample}(d), showing that it is sensitive to the transition in the collective region. The points in the collective region where a transition occurs are given by $\delta C=0$, indicating that the bulk of the ground state changes from one sub-space to the other.

Another criterion that we have proposed~\cite{cordero21, lopez-pena21} in order to find transitions not detectable through the Ehrenfest classification, is to use the Bures distance in the total product space of $n$-level atoms and $\ell$-mode radiation field, defined by~\cite{bures69, uhlmann76}
\begin{equation}
D_B=\sqrt{2}\sqrt{1-|\bra \vec{\alpha},\vec{\gamma}| \vec{\alpha}',\vec{\gamma}'\ket|^2}\,,
\end{equation}
for states
\begin{equation}
\bra \vec{\alpha},\vec{\gamma}| \vec{\alpha}',\vec{\gamma}'\ket = e^{-(|\vec{\alpha}|^2+|\vec{\alpha}'|^2 - 2 \vec{\alpha}^* \cdot \vec{\alpha}')/2} \left( \frac{\vec{\gamma}^*\cdot\vec{\gamma}'}{||\vec{\gamma}|| \, ||\vec{\gamma}'||}\right)^{N_a}\,,
\end{equation}
and maximize it for neighboring states.
As a general procedure, one selects various points around a circumference of radius $\varepsilon$ about each point $p$ in parameter space, in order to find the state with maximum distance to $p$ (cf.~\cite{cordero21, lopez-pena21} for details). In our case, it was sufficient to calculate it for four points about each $p$ in order to get a qualitative behavior of the surface of maximum Bures distance.

\begin{figure*}
\begin{center}
\includegraphics[width=0.48\linewidth]{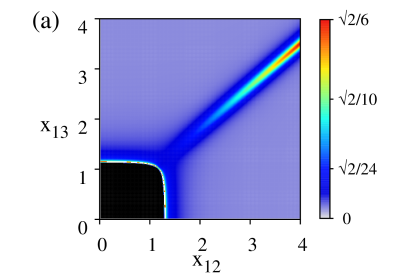}
\includegraphics[width=0.48\linewidth]{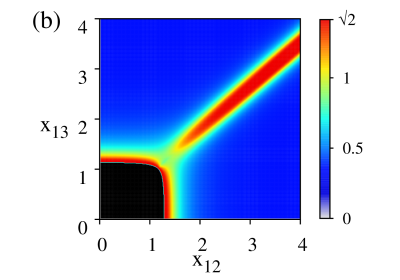}
\caption{(Color online) Surface of maximum Bures distance between neighboring states for (a) $N_a=5$  and (b) $N_a=5000$ particles, in the atomic $V$ configuration. The separatrix within the collective region, which defies an Ehrenfest-type classification, is clearly noticeable. Parameters are the same as in Fig.~\ref{f.Vexample}.}
\label{bures}
\end{center}
\end{figure*}

Figure~\ref{bures} shows, for $N_a=5$ (a) and for $N_a=5000$ (b), the surface of maximum Bures distance between neighboring states. Note that the transition within the collective regions stands out, and for $N_a=5000$ we reach the maximum distance of $\sqrt{2}$ for variational states in the thermodynamic limit. We may refer to this as a transition of the kind {\em continuous unstable}, in the sense that this transition tends to a first order one in the limit $g \to 0$.
%
%

\begin{figure*}
\begin{center}
\includegraphics[width=0.48\linewidth]{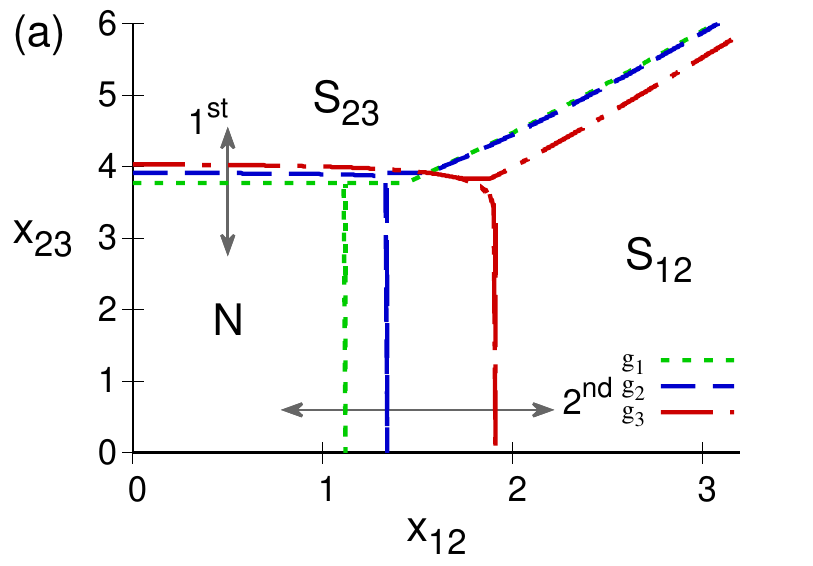}
\includegraphics[width=0.48\linewidth]{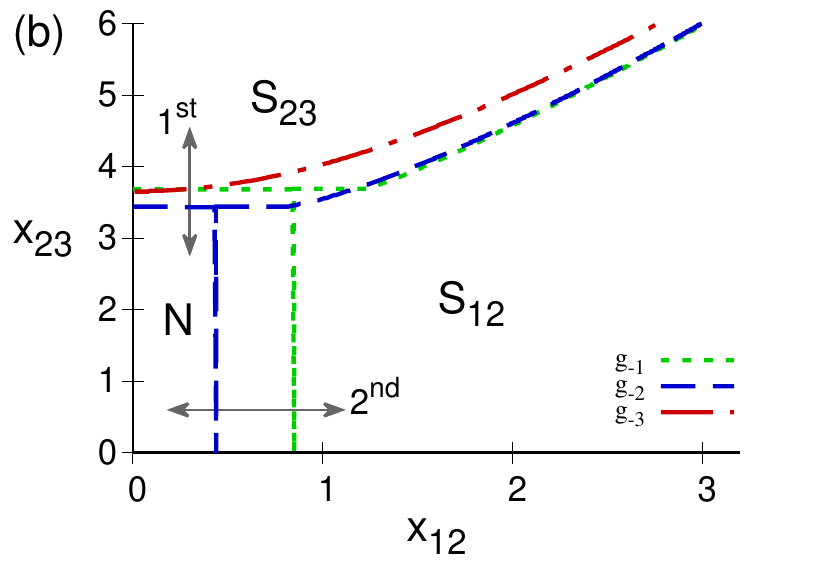}
\caption{(colour online) Separatrices for the $\Xi$-configuration shown as a function of the dimensionless matter-field dipolar strength $x_{jk}$, for values of atomic dipole-dipole strength $g_{\pm1}$ (dotted line), $g_{\pm2}$ (dashed line) and $g_{\pm3}$ (dot-dash line). Figure (a) shows the repulsive, figure (b) the attractive case. Parameters used are discussed in text.}\label{f.sepX}
\end{center}
\end{figure*}
%

\begin{figure*}
\begin{center}
\includegraphics[width=0.48\linewidth]{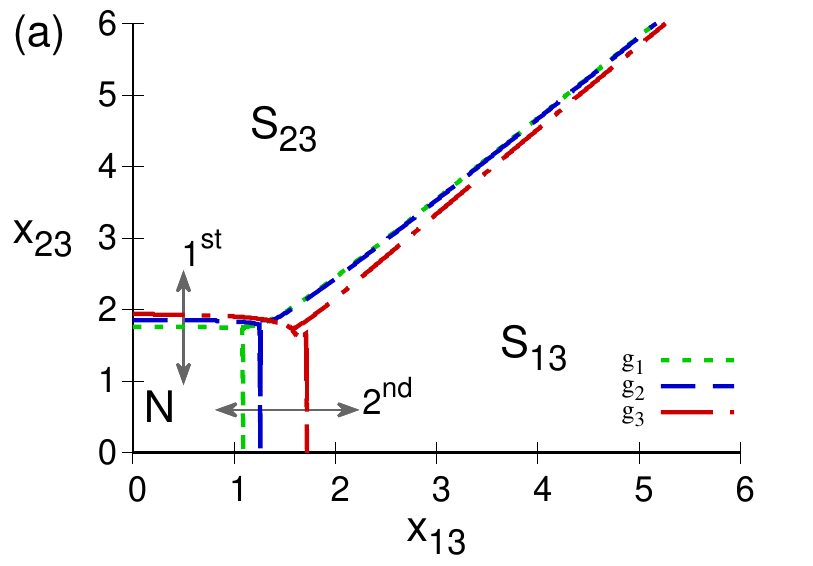}
\includegraphics[width=0.48\linewidth]{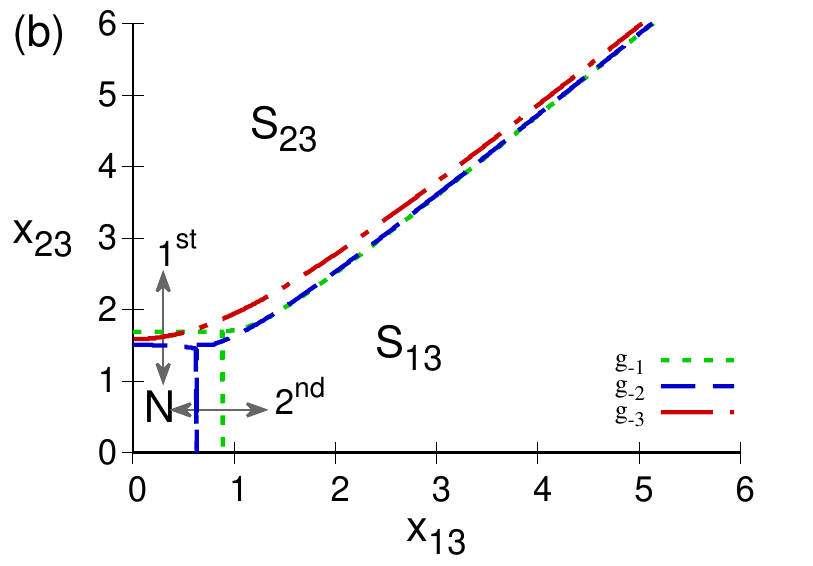}
\caption{(colour online) Separatrices for the $\Lambda$-configuration shown as a function of the dimensionless matter-field dipolar strength $x_{jk}$, for values of atomic dipole-dipole strength $g_{\pm1}$ (dotted line), $g_{\pm2}$ (dashed line) and $g_{\pm3}$ (dot-dash line). Figure (a) shows the repulsive, figure (b) the attractive case. Parameters used are discussed in text.}\label{f.sepL}
\end{center}
\end{figure*}
%

\begin{figure*}
\begin{center}
\includegraphics[width=0.48\linewidth]{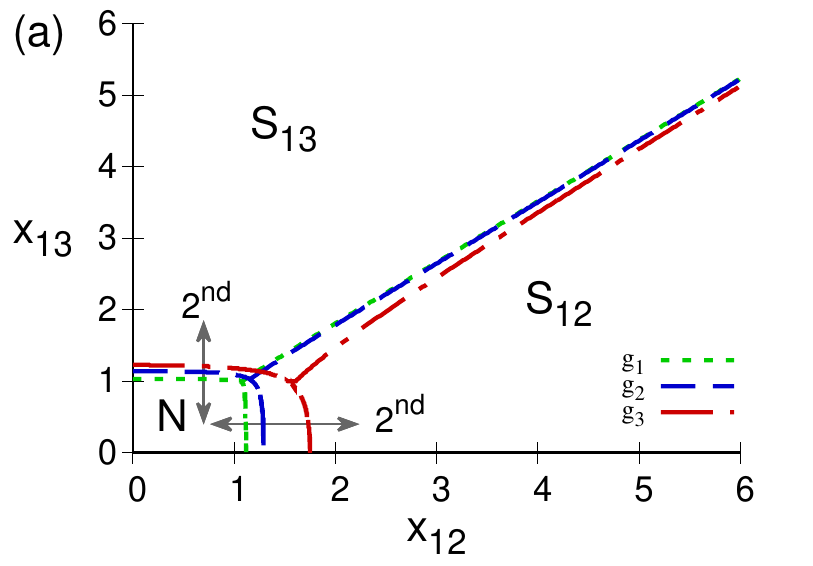}
\includegraphics[width=0.48\linewidth]{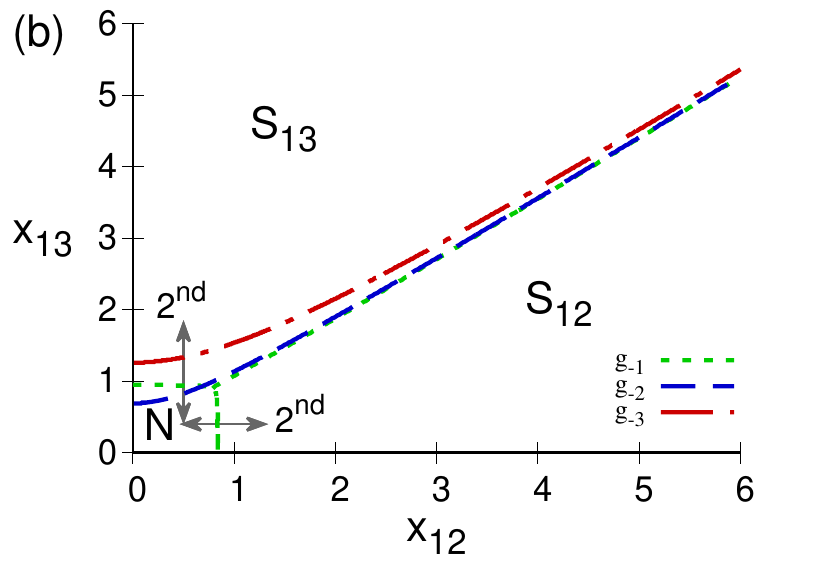}
\caption{(colour online) Separatrices for the $V$-configuration shown as a function of the dimensionless matter-field dipolar strength $x_{jk}$, for values of atomic dipole-dipole strength $g_{\pm1}$ (dotted line), $g_{\pm2}$ (dashed line) and $g_{\pm3}$ (dot-dash line). Figure (a) shows the repulsive, figure (b) the attractive case. Parameters used are discussed in text.}\label{f.sepV}
\end{center}
\end{figure*}

Figure~\ref{f.sepX} shows the separatrix for the atomic $\Xi$-configuration, in the case of a repulsive dipole-dipole interaction~\ref{f.sepX}(a), and in the case of an attractive one attractive~\ref{f.sepX}(b). One notes that, in the repulsive case, the normal region $N$ grows as the dipole-dipole interaction grows. The regions where the bulk of the ground state is dominated by the basis of the subsystem $S_{12}$ or $S_{23}$ are also indicated. The order of the phase transitions are marked: a first order transition for $N \leftrightarrow S_{23}$ and second order transition for $N\leftrightarrow S_{12}$. In the attractive case, Fig.~\ref{f.sepX}(b), the normal region decreases in size as $g$ increases in magnitude, and it in fact vanishes for the value of $g_{-3}$ where only the regions $S_{12}$ and $S_{23}$ subsist; for $g_{-1}$ and $g_{-2}$ the order of the phase transitions is the same as in the repulsive case.  

A similar behavior occurs for the $\Lambda$-configuration, Fig.~\ref{f.sepL}, where the subregions in the collective regime are $S_{13}$ and $S_{23}$.

Figure~\ref{f.sepV} shows the situation for atoms in the $V$-configuration. In the repulsive case, Fig~\ref{f.sepV}(a), a normal region exists in all the circumstances, and the transitions from the normal to the collective region are of second order. For the attractive case, Fig~\ref{f.sepV}(b), in the case $g_{-1}$ (dotted line) a normal region exists and we have a second order transition. In the strong attractive cases of $g_{-2}$ and $g_{-3}$ we only have the collective regions $S_{12}$ and $S_{13}$. 

\section{Conclusions}
\label{conclusions}

We have established the general atomic dipole-dipole interaction Hamiltonian for a system of $n$-level atoms interacting with $\ell$-modes of electromagnetic radiation in a cavity, together with the associated energy surface, which allows to determine the variational ground state (see expressions~(\ref{eq.Ev}), (\ref{eq-dip-dip}), and (\ref{eq.rsc1})). For  $2$- and $3$-level atomic configurations, we have found that for attractive (repulsive) atomic dipole-dipole interactions the normal region decreases (increases) in size. The quantum phase diagrams, together with the corresponding order of the transitions, have also been determined. For a finite or infinite number of atoms, the surface of maximum Bures distance is able to detect the transitions between the collective regions where the Ehrenfest criterion fails (see Fig.~\ref{bures}). In other words, we find that, in cases where the Ehrenfest criterion for the phase transitions does not give information, a criterion based on the maximum probability for prohibited transitions comes to the rescue. We have also proved that the quantum phase diagrams continue being dominated by monochromatic regions as it is the case for noninteracting atoms, at least for real induced electric dipolar moments.

Phase diagrams for $2$- and $3$-level atoms interacting with an external radiation field have been studied, for all the possible atomic configurations. It is seen that the atomic dipole-dipole interaction is minuscule compared with the dipolar matter-field interaction, so the atomic dipole-dipole coupling has been exaggerated in order to see its consequences. (The unnaturally large values for this coupling, taken so that its effects may be appreciated, must be scaled down accordingly when studying actual realistic systems.) Although small, energy transfer between the particles (atoms, molecules) is one of the important consequences of this interaction, as is evident in the Van der Waals forces between induced dipoles. The formation of optical lattices, and the many-body effects in systems such as atomic clocks, are also some of its consequences~\cite{lahaye09}.

The separatrices dividing normal from collective superradiant regions have been calculated and classified according to the Ehrenfest classification. However, there are separatrices present within the collective regimes, marking transitions between regions where one or another mode of the radiation field dominates the bulk of the ground state, which defy the Ehrenfest classification. In these cases, we have proposed two methods to detect, calculate, and classify them, one based on the second Casimir operator and another one using the surface of maximum Bures distance between neighboring states.

\appendix
\section{Matter collective operators}
\label{ap.collective}
Let $\op{A}_{pq}^{(\gamma)}$  denote the matter operator of the $\gamma$th atom of $n$-levels, which promotes the atom from level $\omega_q$ to level $\omega_p$.  Note that $\op{A}_{qp}^{(\gamma)}={\op{A}_{pq}^{(\gamma)}}^\dag$. For each atom $\gamma$ these operators obey the unitary algebra $u_\gamma(n)$ in $n$ dimensions (for $n$-level atoms), i.e.,
\begin{eqnarray}
&& \sum_{q=1}^n \op{A}_{qq}^{(\gamma)} = \op{1}_\gamma\,, \\[3mm]
&& \left[\op{A}_{pq}^{(\gamma)},\op{A}_{rs}^{(\gamma')} \right] = \delta_{\gamma\gamma'}\left(\delta_{qr} \op{A}_{ps}^{(\gamma)}- \delta_{ps}\op{A}_{rq}^{(\gamma)}\right)\,,\label{comm.gamma}
\end{eqnarray}
with $\op{1}_\gamma$ the identity operator in the subspace $\gamma$. Also note that, for a single atom, we have
\begin{equation}\label{eq.ApqArs.g}
\op{A}_{pq}^{(\gamma)}\op{A}_{rs}^{(\gamma)} = \delta_{qr}\,\op{A}_{ps}^{(\gamma)}\,.
\end{equation}

For $N_a$ identical atoms, the collective matter operator is defined as 
\begin{equation}
\op{A}_{pq}:={\sum_{\gamma=1}^{N_a}}  \op{A}_{pq}^{(\gamma)}\,,
\end{equation}
and note that the sum over $\gamma$ does not preserve the structure of the each subspace. 
By simple inspection, one may prove easily the follow relationships for the collective operators:
\begin{equation}
 \op{A}_{qp}=\op{A}_{pq}^\dag\,,
 \label{eq.qppq}
\end{equation}
\begin{equation}
 \sum_{q=1}^{n} \op{A}_{qq} = \sum_{\gamma=1}^{N_a} \op{1}_{\gamma}:=N_a\,\op{1}\,,  \label{eq.na}
\end{equation}
\begin{equation}
 \sum_{j,k=1}^n \op{A}_{kj}\op{A}_{jk} = N_a\,(N_a+n-1)\, ,
 \label{casimir2}
\end{equation}
\begin{equation}
 \left[\op{A}_{pq},\op{A}_{rs} \right] = \delta_{qr} \op{A}_{ps}-   \delta_{ps}\op{A}_{rq}\,.
 \label{eq.comm}
\end{equation}
Equations (\ref{eq.na}) and (\ref{casimir2}) are the first and second order Casimir operators; equation (\ref{eq.comm}) shows that the operators $\op{A}_{pq}$ obey a unitary algebra in $n$ dimensions, $U(n):= \oplus_{\gamma=1}^{N_a} u_\gamma(n)$. The weight operators are $\op{A}_{pp}$ which give the number of particles in each atomic level $\omega_p$, i.e., for an uncoupled state $|\psi\ket$ one has  $\op{A}_{pp}|\psi\ket = n_p|\psi\ket$ with $n_p$ the atomic population, while the operator $\op{A}_{pq}$ (with $p\neq q$) promotes the transition of one atom from the level $\omega_q$ to the level $\omega_p$; this is clear from (\ref{eq.comm}) since, for the uncoupled state $|\psi\ket$ with atomic populations $n_p$ and $n_q$ in the atomic levels $\omega_p$ and $\omega_q$ respectively (i.e.,  $\op{A}_{pp}|\psi\ket = n_p|\psi\ket$ and $\op{A}_{qq}|\psi\ket = n_q|\psi\ket$), after applying $\op{A}_{pq}|\psi\ket=|\psi'\ket$ one has $\op{A}_{pp}|\psi'\ket = (n_p+1) |\psi'\ket$ and $\op{A}_{qq}|\psi'\ket = (n_q-1) |\psi'\ket$, while the other atomic populations are preserved.

In similar fashion to equation~(\ref{eq.ApqArs.g}), and using~(\ref{comm.gamma}), one finds
\begin{equation}\label{eq.ApqArs}
\op{A}_{pq}\op{A}_{rs}= \op{A}_{ps}\left(\op{A}_{rq}+\delta_{rq}\right) - \delta_{rs}\op{A}_{pq} + \op{O}_{pqrs}\,,
\end{equation}
where
\begin{equation}\label{eq.O}
 \op{O}_{pqrs}=\sum_{\gamma\neq\gamma'}^{N_a} \left(\op{A}_{pq}^{(\gamma)}\op{A}_{rs}^{(\gamma')}-\op{A}_{ps}^{(\gamma)}\op{A}_{rq}^{(\gamma')}\right)\,.
\end{equation}

It is straightforward to show the relationships $\op{O}_{pqrs}=-\op{O}_{psrq}$, $\op{O}_{pqrs}=\op{O}_{rspq}$ and $\op{O}_{pqrq}=0$.  Also, for totally symmetric particles, where one may use the bosonic representation of the collective operators, one has the identity $\op{O}_{pqrs}=0$.

We define the {\em oslash-product} $\oslash$ as the product of matter collective operators  without self-interaction
\begin{equation}\label{eq.ApqSArs}
\op{A}_{pq}  \oslash \op{A}_{rs} := \sum_{\gamma\neq\gamma'}^{N_a} \op{A}_{pq}^{(\gamma)}\,\op{A}_{rs}^{(\gamma')}= \op{A}_{pq}\,\op{A}_{rs} -\delta_{qr}\op{A}_{ps}\,.
\end{equation}
Notice that $\op{A}_{pq} \oslash \op{A}_{rs} = \op{A}_{rs} \oslash \op{A}_{pq}$ and also $\op{O}_{pqrs} = \op{A}_{pq} \oslash \op{A}_{rs} -\op{A}_{ps} \oslash \op{A}_{rq}$, so that by replacing~(\ref{eq.ApqSArs}) into equation~(\ref{eq.ApqArs}) the latter is satisfied trivially.

\begin{widetext}
\section{Dipole-Dipole Operator}\label{ap.dd}

The atomic dipole-dipole interaction is written as in Eq.~(\ref{eq.opWa00}) 
\begin{equation}\label{eq.opWa}
\op{H}_{dd}=\frac{1}{2(N_a-1)}\sum_{j\neq k}^n\sum_{l \neq m}^n g_{jklm}\op{A}_{jk}\oslash\op{A}_{lm}\,.
\end{equation}
Taking into account the symmetries between the indices of $g_{jklm}$, and the possible transitions shown in table~\ref{t.1}, we need only to replace the {\em oslash} product in~(\ref{eq.ApqSArs}) for the dipole-dipole operator~(\ref{eq.opWa}) to read
%
\begin{eqnarray}
\op{H}_{dd}&=&\frac{1}{2(N_a-1)}\sum_{j\neq k}^n\left[ g_{jkjk}\op{A}_{jk}\op{A}_{jk} +  g_{jkkj}(\op{A}_{jk}\op{A}_{kj}-\op{A}_{jj})\right] 
 + \frac{1}{2(N_a-1)}\sum_{j\neq k\neq l}^n \left[g_{jkjl}\op{A}_{jk}\op{A}_{jl} + g_{jklj}\op{A}_{jk}\op{A}_{lj} \right.\nonumber\\
&& + \left. g_{jklk}\op{A}_{jk}\op{A}_{lk} +g_{jkkl}(\op{A}_{jk}\op{A}_{kl} - \op{A}_{jl}) \right]
 +\frac{1}{2(N_a-1)}\sum_{j\neq k\neq l \neq m}^n g_{jklm}\op{A}_{jk}\op{A}_{lm}\,,
\label{B2}
\end{eqnarray}
%
where the first line refers to single dipole-dipole interactions, the second line to the interaction between dipoles which share an atomic level, and the third line to separate dipoles not sharing atomic levels. 

We may rewrite the atomic dipole-dipole operator as
\begin{eqnarray}\label{eq.W}
\op{H}_{dd} &=& \frac{1}{2!}\sum_{j\neq k}^n \op{W}_{jk}^{2-{\rm levels}} + \frac{1}{2!}\sum_{j\neq k\neq l}^n \op{W}_{jlk}^{3-{\rm levels}} 
 +\frac{1}{4!} \sum_{j\neq k\neq l \neq m}^n  \op{W}_{jklm}^{4-{\rm levels}}\,. 
\end{eqnarray}
with
%
\begin{eqnarray}
\op{W}_{jk}^{2-{\rm levels}}&=& \frac{1}{2(N_a-1)}\left(g_{jkjk} \op{A}_{jk}\op{A}_{jk} + g_{kjkj} \op{A}_{kj}\op{A}_{kj}\right) 
 +  \frac{1}{N_a-1}g_{jkkj}(\op{A}_{jk}\op{A}_{kj}-\op{A}_{jj})\,, \label{eq.W2levels} \\[3mm]
\op{W}_{jkl}^{3-{\rm levels}}  &=&\frac{1}{2(N_a-1)}\left( g_{jklk}\{\op{A}_{jk},\op{A}_{lk}\} + g_{kjkl}\{\op{A}_{kj},\op{A}_{kl}\} \right)\nonumber \\
&& + \frac{1}{N_a-1} \left[g_{jkkl}(\op{A}_{jk}\op{A}_{kl}-\op{A}_{jl}) +g_{kjlk}(\op{A}_{kj}\op{A}_{lk}-\op{A}_{kk})\right]\,,  \label{eq.W3levels}\\[3mm]
\op{W}_{jklm}^{4-{\rm levels}} &=&  \frac{1}{N_a-1}\left( g_{j k l m} \op{A}_{jk}\op{A}_{lm}+
 g_{j k m l} \op{A}_{jk}\op{A}_{ml}\right.  +
 g_{j l k m} \op{A}_{jl}\op{A}_{km}+
 g_{j l m k} \op{A}_{jl}\op{A}_{mk}\nonumber \\ && +
 g_{j m k l} \op{A}_{jm}\op{A}_{kl}+
 g_{j m l k} \op{A}_{jm}\op{A}_{lk} +
 g_{k j l m} \op{A}_{kj}\op{A}_{lm} +
 g_{k j m l} \op{A}_{kj}\op{A}_{ml}\nonumber \\ && +
 g_{k l m j} \op{A}_{kl}\op{A}_{mj}+
 g_{k m l j} \op{A}_{km}\op{A}_{lj}  + \left.
 g_{l j m k} \op{A}_{lj}\op{A}_{mk}+
 g_{l k m j} \op{A}_{lk}\op{A}_{mj}\right)\,,\qquad
\end{eqnarray}
\end{widetext}
where $\{\op{A}_{jk},\op{A}_{lm}\} = \op{A}_{jk}\op{A}_{lm}+\op{A}_{lm}\op{A}_{jk}$ is the anti-commutator of $\op{A}_{jk}$ and $\op{A}_{lm}$. The factor $1/p!\,,\ (p=2,4)$ in~Eq.~(\ref{eq.W}) eliminates the double summation, because $\op{W}_{jk}^{2-{\rm levels}}=\op{W}_{kj}^{2-{\rm levels}}$, $\op{W}_{jkl}^{3-{\rm levels}}=\op{W}_{lkj}^{3-{\rm levels}}$ and $\op{W}_{jklm}^{4-{\rm levels}} =\op{W}_{\sigma(jklm)}^{4-{\rm levels}}$, with $\sigma(jklm)$ a permutation of the indices $(jklm)$. 

The contribution to the atomic dipole-dipole interaction given in~(\ref{eq.W2levels}) corresponds to transitions $\omega_j\rightleftharpoons \omega _k$ similar to a $2$-level atom, while the contribution in~(\ref{eq.W3levels}) promotes the atomic transitions  $\omega_j\rightleftharpoons \omega _l$ via an intermediate atomic level $\omega_k$; here, the direct dipolar transition $\omega_j\rightleftharpoons \omega _l$ is prohibited. This contribution $\op{W}_{jkl}^{3-{\rm levels}}$ appears for $n$-level atoms with $n\geq 3$. The last term in Eq.~(\ref{eq.W}) promotes transitions between two unconnected permitted dipolar transitions $\omega_j\rightleftharpoons \omega _k$ and $\omega_l\rightleftharpoons \omega _m$, and is present for $n$-level atoms with $n\geq 4$.

As an example, for $2$-level atoms the dipole-dipole interaction reads
\begin{eqnarray}
\op{H}_{dd} &=& \op{W}_{12}^{2-{\rm levels}} \,,
\end{eqnarray}
while for $3$-level atoms one finds the following for each configuration:
\begin{itemize}
\item $\Xi$-configuration with prohibited dipolar transition $\omega_1\rightleftharpoons \omega_3$ ($\vec{d}_{13}=\vec{0}$)
\begin{equation}
\op{H}_{dd}^{(\Xi)} = \op{W}_{12}^{2-{\rm levels}} +  \op{W}_{23}^{2-{\rm levels}}  + \op{W}_{123}^{3-{\rm levels}} \,.
\end{equation}

\item $\Lambda$-configuration with prohibited dipolar transition $\omega_1\rightleftharpoons \omega_2$ ($\vec{d}_{12}=\vec{0}$)
\begin{equation}
\op{H}_{dd}^{(\Lambda)} =  \op{W}_{13}^{2-{\rm levels}}  + \op{W}_{23}^{2-{\rm levels}} + \op{W}_{132}^{3-{\rm levels}} \,,
\end{equation}

\item $V$-configuration with prohibited dipolar transition $\omega_2\rightleftharpoons \omega_3$ ($\vec{d}_{23}=\vec{0}$)
\begin{eqnarray}
\op{H}_{dd}^{(V)}&=&  \op{W}_{12}^{2-{\rm levels}}  +  \op{W}_{13}^{2-{\rm levels}}  + \op{W}_{213}^{3-{\rm levels}}\,.
\end{eqnarray}
\end{itemize}

Finally, we evaluate the dipole-dipole operator for two $4$-level atomic configurations. In the particular case of the $\lambda$-configuration, with prohibited transitions $\vec{d}_{12}=\vec{d}_{14}=\vec{d}_{24}=\vec{0}$, the dipole-dipole operator reduces to
\begin{eqnarray}
\op{H}_{dd}^{(\lambda)} &=& \op{W}_{13}^{2-{\rm levels}} +\op{W}_{23}^{2-{\rm levels}} +\op{W}_{34}^{2-{\rm levels}} \nonumber \\ 
&& + \op{W}_{134}^{3-{\rm levels}} + \op{W}_{234}^{3-{\rm levels}} + \op{W}_{132}^{3-{\rm levels}}\,;
\end{eqnarray}
notice that in this case we have no contribution of the form $\op{W}_{1234}^{4-{\rm levels}}$ because all atomic levels are connected via the atomic level $\omega_3$. 

On the other hand, for atoms in the $\largelozenge$-configuration the prohibited dipolar transitions are $\vec{d}_{14}=\vec{d}_{23}=\vec{0}$ and, since this atomic configuration has isolated dipoles, the total dipole-dipole operator has a non-zero contribution from $\op{W}_{1234}^{4-{\rm levels}}$\,:
\begin{eqnarray}
\op{H}_{dd}^{(\medlozenge)} &=& \op{W}_{12}^{2-{\rm levels}} +\op{W}_{13}^{2-{\rm levels}} +\op{W}_{24}^{2-{\rm levels}}  +\op{W}_{34}^{2-{\rm levels}}
\nonumber \\ && +
\op{W}_{124}^{3-{\rm levels}} + \op{W}_{134}^{3-{\rm levels}} + \op{W}_{213}^{3-{\rm levels}} + \op{W}_{243}^{3-{\rm levels}}
\nonumber \\ && +
\op{W}_{1234}^{4-{\rm levels}}\,.
\end{eqnarray}

\end{document}